# Examining the performance of refractory conductive ceramics as plasmonic materials: a theoretical approach

Mukesh Kumar,[1,2,*] Naoto Umezawa,[1,2] Satoshi Ishii,[2,3] and Tadaaki Nagao[2,3]

[1]*Environmental Remediation Materials Unit, National Institute for Materials Science, Ibaraki 305-0044, Japan*
[2]*CREST, Japan Science and Technology Agency, 4-1-8 Honcho, Kawaguchi, Saitama, 332-0012, Japan*
[3]*International Center for Materials Nanoarchitectonics (MANA), National Institute for Materials Science (NIMS), Tsukuba 305-0044, Japan*
*\*mkgarg79@gmail.com;Kumar.Mukesh@nims.go.jp*

---

**Abstract**

The main aim of this study is to scrutinize promising plasmonic materials by understanding their electronic structure and correlating them to the optical properties of selected refractory materials. For this purpose, the electronic and optical properties of the conductive ceramics TiC, ZrC, HfC, TaC, WC, TiN, ZrN, HfN, TaN, and WN are studied systematically by means of first-principles density functional theory. A full *ab initio* procedure to calculate plasma frequency from the electronic band structure is discussed. The dielectric functions are calculated by including electronic interband and intraband transitions. Our calculations confirm that transition metal nitrides, such as TiN, ZrN, and HfN, are the strongest candidates, exhibiting performance comparable to that of conventional noble metals in the visible to the near-infrared regions. On the other hand, carbides are not suitable for plasmonic applications because they show very large losses in the same regions. From our calculated dielectric functions, the scattering and absorption efficiencies of nanoparticles made of these refractory materials are evaluated. It is revealed that TiN and TaC are the best candidate materials for applications in photothermal energy conversion over a broad spectral region. Furthermore, quality factors for localized surface plasmon resonance and surface plasmon polaritons are calculated to compare quantitative performances, and ZrN and HfN are found to be comparable to conventional plasmonic metals such as silver and gold.

---

## I. INTRODUCTION

Metal nanoplasmonics has become one of the central research topics in nanotechnology and has opened up a wide range of applications in a variety of fields, such as medical therapeutics/diagnostics, photovoltaic devices, optical metamaterials, subwavelength communication devices, as well as lasers.[1-10] Many applications of plasmonics require materials with high electrical conductivity. This implies that materials with an abundance of free electrons (i.e., large negative real permittivity) and low electron scattering rate (i.e., low interband losses) are preferable. Interband losses generally arise from interband electronic transitions (imaginary permittivity), transitions from the Fermi surface to the next higher empty conduction band (CB) or from a lower lying filled valence band (VB) to the Fermi level ($E_F$) or higher levels. Because metals tend to have high electrical conductivity and large negative permittivity, they have traditionally been the materials of choice for plasmonics.[10] Among metals, Au and Ag are commonly used in plasmonic devices because of their excellent properties and chemical stability in many environments.[10] Apart from Au and Ag, recent studies have shown that other metals such as aluminum, gallium, platinum, and palladium are also useful for plasmonic applications in the ultraviolet (UV) region.[11-14] In addition to using metals as single elements, alloying noble metals with other metals, such as platinum, palladium, and iron, can enhance plasmonic catalytic activities and plasmonic absorptions.[15-17] However, plasmonic materials are not restricted to elemental metals or alloys. There is growing interest in alternative plasmonic materials such as semiconductors, transparent conducting oxides (TCOs), and intermetallics.[18] Compound refractory materials are another class of alternative plasmonic materials that have high melting point temperatures (> 1500 ºC) and excellent chemical stability.[19] In this context, TiN and ZrN have already received great attention recently and have been considered as promising candidates for plasmonic applications at optical frequencies and at high temperatures.[20,21]

In the field of computational materials science, density functional theory (DFT) based on the Kohn-Sham formulation has become one of the important computational methods to model the ground-state properties of materials from first-principles. Many studies have shown that electronic band structure and optical properties of inorganic materials can be well described from DFT by using appropriate approximations.[22,23] Inspired by those studies, investigators in the field of plasmonics have made several attempts using first-principles DFT to understand materials properties, especially dielectric functions of metals, doped TCOs, and intermetallic or other compounds.[24-31] However, to the best of our knowledge, no such attempts have been made for refractory materials. Therefore, in this article, in addition to TiN and ZrN, we discuss the suitability of various refractory materials, in particular, TiC, ZrC, HfC, TaC, WC, HfN, TaN, and WN, for plasmonic application by studying their electronic and optical properties with the help of first-principles DFT. In addition, classical electromagnetic (EM) calculations were also carried out to study the scattering and absorption efficiencies of these refractory materials in nanosphere form. Performance of each refractory material is evaluated based on various quality (Q) factors defined for different classes of plasmonic devices. At the end of each section, a comparative study of selected refractory materials and conventional noble metals, such as Au and Ag, is performed.

This paper is organized as follows: In Sec. II we give details of the theoretical methods (DFT and EM) used in this study. In Sec III we present the results and discussion in two parts: part A, analysis of bulk properties, and part B, analysis of properties of nanostructures.

**II. MODEL AND METHODS**

**First-principles calculation**

For a description of the dielectric function from electronic band structure calculations, we apply the independent particle approximation (IPA). IPA is a successful approach for examining the optical permittivity of many metallic systems.[26,31] The dielectric function $\varepsilon(\omega) = \varepsilon_1(\omega) + i\varepsilon_2(\omega)$ (where $\varepsilon_1$ is the real part, and $\varepsilon_2$ is the imaginary part) of metals (semimetals) in the low frequency region can be described by taking into account two important processes, interband [$\varepsilon^{interband}(\omega)$] and intraband [$\varepsilon^{intraband}(\omega)$] transitions. (The dielectric function for intraband transitions, $\varepsilon^{intraband}(\omega)$, is also known as the dielectric function of a free electron gas.) Hence, the total dielectric function is the sum of the dielectric functions for inter- and intraband transitions:

$$\varepsilon(\omega) = \varepsilon^{interband}(\omega) + \varepsilon^{intraband}(\omega) \quad (1)$$

The contribution of interband excitations to the dielectric function, $\varepsilon^{interband}(\omega)$, was obtained from electronic band structure calculation by evaluating matrix elements for direct electronic transitions between occupied and unoccupied electronic states following the method discussed by Kumar et al:[32]

$$\varepsilon^{interband}(\omega) = \varepsilon_1^{interband}(\omega) + i\varepsilon_2^{interband}(\omega) \quad (2)$$

Here, the imaginary part $\varepsilon_2^{interband}(\omega)$ was calculated from the joint density of states (DOS) and the optical momentum matrix in the long wave length limit, $q = |\mathbf{k}' - \mathbf{k}| \to 0$, using the following equation:[33]

$$\varepsilon_2^{interband}(\omega) = \frac{4\pi^2 e^2}{\Omega} \lim_{q \to 0} \frac{1}{q^2} \sum_{c,v,\mathbf{k}} 2w_\mathbf{k}\, \delta(E_{c\mathbf{k}} - E_{v\mathbf{k}} - \omega) \times \langle u_{c\mathbf{k}+e_x q}|u_{v\mathbf{k}}\rangle \langle u_{c\mathbf{k}+e_y q}|u_{v\mathbf{k}}\rangle, \quad (3)$$

where $e$ is the electron charge, $\Omega$ is the primitive cell volume, $w_\mathbf{k}$ is the weight of the **k**-points, and $\mathbf{e}_x$ and $\mathbf{e}_y$ are the unit vectors for the three Cartesian directions. The indices $c$ and $v$ refer to the CB and VB states respectively. The parameter $E_{j\mathbf{k}}$ is the single-electron energy state of band $j$ at wave vector **k**, $u_{j\mathbf{k}}$ is the periodic part of the Bloch wave function corresponding to the eigenvalue $E_{i\mathbf{k}}$ ($i = c, v$) and $\delta$ is the delta function, which depends upon the method used for calculation.

The real part $\varepsilon_1^{interband}(\omega)$ was obtained from the Kramers-Kronig transformation using the following equation:[33]

$$\varepsilon_1^{interband}(\omega) = 1 + \frac{2}{\pi} P \int_0^\infty \frac{\varepsilon_2^{int}(\omega')\omega'}{\omega'^2 - \omega^2 + i\eta} d\omega' \quad (4)$$

where $P$ is the principal value and $\eta$ is an infinitesimal.

The dielectric function for intraband transitions, $\varepsilon^{intraband}(\omega)$, was obtained by approximating it with the

Drude expression for a given plasma frequency ($\omega_p$) and damping constant ($\gamma$),[34,35]

$$\varepsilon(\omega) = 1 - \frac{\omega_p^2}{w^2 + i\gamma\omega}, \quad (5)$$

The real ($\varepsilon_1^{intraband}$) and imaginary ($\varepsilon_2^{intraband}$) components of this dielectric function are given by

$$\varepsilon_1^{intraband}(\omega) = 1 - \frac{\omega_p^2 \tau^2}{1 + w^2 \tau^2}, \quad (6)$$

$$\varepsilon_2^{intraband}(\omega) = \frac{\omega_p^2 \tau}{\omega(1 + \omega^2 \tau^2)}, \quad (7)$$

where $\tau$ ($\gamma = 1/\tau$) represents a relaxation time obtained from a higher-order calculation while $\omega_p$ is the plasma frequency calculated from the band structure as follows:[26]

$$(8)$$

Once $\omega_p$ and $\gamma$ are obtained, the real and imaginary components of the intraband part of the dielectric function were obtained from equations 6 and 7.

All these calculations were performed using DFT as implemented in the plane-wave Vienna ab initio Simulation Package.[36] The generalized gradient approximation (GGA) of Perdew-Burke-Ernzerhof (PBE) was used for the exchange-correlation potential.[37] The lattice constants were relaxed by minimizing the total energy to an accuracy of 0.1 meV, and the atomic positions were relaxed with an accuracy of 0.01 eV/Å for the forces on each atom. An energy cutoff of 500 eV and a Brillouin integration grid built using the Monkhorst-pack scheme with 21×21×21 points were used for the self-consistent calculations.[38] However, for optical-properties calculations, a grid of size 69×69×69, resulting approximately 12,000 **k**-points, along with a sufficient number of conduction bands was used. Increasing the number of **k**-points beyond this was tested and was confirmed to exhibit any substantial change in the spectral details. The plasma frequency was well converged to within 1-2% of the experimental values. The most stable rocksalt structure (Fm-3m: space group No. 225) of these refractory materials was considered for calculations in this study.[39] It should be noted that the stable structures of WC and WN are different.[39] However, to maintain validity of comparisons, here we studied the rocksalt structure only.

**Classical electromagnetic calculations**

Based on the theory of Mie, which is the analytic solution for a single sphere illuminated by a plane wave, one can obtain the scattering ($Q_{sca}$) and absorption ($Q_{abs}$) cross section of a spherical particle.[40]

## III. RESULTS AND DISCUSSION

### Part A: Bulk properties calculations

In this section we discuss the electronic and optical properties of the ten chosen refractory materials and then compare the performance of these materials with that of Au and Ag in terms of permittivity and various other quality factors.

Our DFT-calculated lattice constants of these materials (shown in Table 1) are in good agreement (within 1-3% error range) with reported experimental and theoretical lattice constants.[39,41]

**Electronic properties of refractory materials**

Figure 1 shows the calculated band structures along high symmetry directions of these crystals.

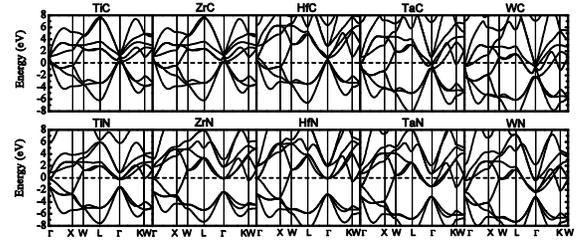

**Figure 1.** Calculated electronic band structure of refractory materials plotted along high symmetry directions of the first Brillion zone (BZ). The black dashed line indicates the Fermi-level, which is set at 0 eV.

As seen in Fig. 1, all of these refractory materials exhibit a metallic electronic structure; $E_F$ (dotted line in each figure) crosses either the upper or lower band. One can see that the overall features for all studied materials are similar except for the position of $E_F$. For carbides, $E_F$ is situated at the top of the lower band (except TaC and WC), while for nitrides, it is located in the middle of the upper band. The different $E_F$ position is attributed to the fact that nitrogen has more valence electrons (five) than carbon (four). A similar trend can be seen in cases of Hf, Ta, and W. As the number of valence electrons increases from HfN (four) to TaN (five) to WN (six), $E_F$ shifts upwards. To understand the contribution of each atomic orbital in the band structures, we studied the orbital-projected DOS as shown in Fig. 2. However, to avoid multiple similar graphs, we show only one representative case, TiC and TiN, for comparison of carbides and nitrides.

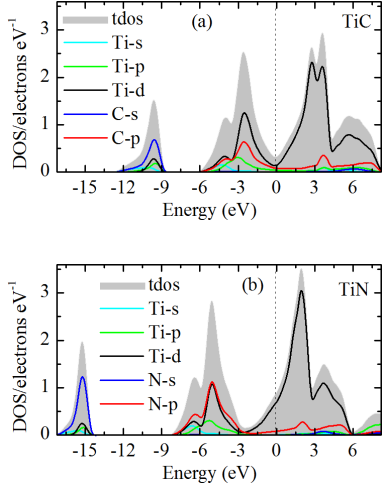

**Figure 2.** DOS of (a) TiC and (b) TiN as representative of carbide and nitride refractory materials.

It is clearly seen that all these compounds consist of two major bands (upper and lower) near $E_F$. The lower band is mainly composed of non-metallic (C or N) $p$ orbitals hybridized with the metallic Ti $p$ and Ti $d$ orbitals (and similarly for other metallic $d$ orbitals). It is reported that this hybridization is the origin of the strong covalent bonding of these refractory materials, which is responsible for their high chemical stability.[41] The half-filled $d$ orbital of metallic Ti dominates the upper band, which features steep dispersion around $E_F$ and is responsible for the high metallicity. Overall, one can expect (C or N) $p \rightarrow$ Ti $d$ or Ti $p \rightarrow$ Ti $d$ interband transitions to contribute to major optical losses. Importantly, the position of the $p$ band with respect to $E_F$ is lower in TiN than in TiC by around 3 eV. This indicates that a higher energy (in the UV region) is required for the interband $p \rightarrow d$ transitions to occur in TiN. On the other hand, the $s$ orbitals of non-metallic species are far below $E_F$ (~ 9 eV for carbides and ~ 15 eV for nitrides) and thus do not contribute to optical losses in the energy range of solar the spectrum. From these observations, one can expect that TiC exhibits a larger loss in the visible region and a nearly comparable or smaller loss in the deep UV region compared to TiN. Similar analysis is valid for the other compounds as well.

**Optical properties of refractory materials**

Figure 3 shows the calculated dielectric functions for various refractory materials, and Table 1 lists several calculated parameters, which are in good agreement with previously reported data. The most distinctive differences are the onset of interband transitions ($\omega_{int}$), the width of the plasmon resonance ($W_P$, corresponding to the value of $\varepsilon_2(\omega)$ when $\varepsilon_1(\omega) = 0$), the cross-over frequency ($\omega_c$, corresponding to the value when $\varepsilon_1(\omega)$ crosses zero), and the plasma frequency ($\omega_p$). The calculated $\omega_p$ values of 7.49 eV for TiN, 8.63 eV for ZrN, and 9.10 eV for HfN are in good agreement with previously reported experimental (theoretical) values of 6.69-8.7 (7.62) eV for TiN, 7.17-8.02 (8.82) eV for ZrN, and 7.32-8.63 eV for HfN.[10,31,42]. However no previous reports are found for other refectory materials listed in Table 1.

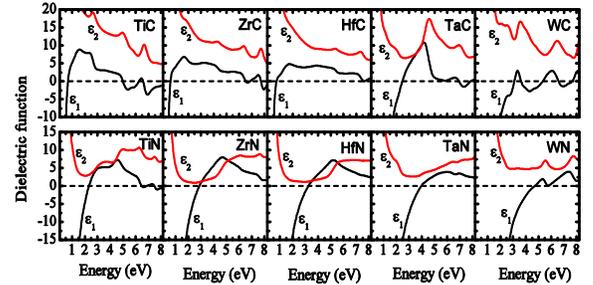

**Figure 3.** Calculated permittivities of various refractory materials by DFT.

It is noticeable that the interband contribution for nitrides is very weak compared to that in carbides in the low energy region up to ~4 eV. The reason is due to the electronic band structures of these materials. As we discussed earlier, the interband $p \rightarrow d$ transition costs more energy in nitrides because the occupied $p$ band, which generally influences $\omega_{int}$ and optical losses, is much lower than that in carbides. This is the reason that the threshold energy for interband transitions is $\geq 2$ eV for nitride compounds compared to that in their carbide counterparts, which is < 1 eV. As the position of $p$ states varies from ZrN to HfN, $\omega_{int}$ increases accordingly (see Table 1) and influences the optical losses.

**Table 1.** Calculated lattice parameter ($a$), plasma frequency ($\omega_p$), cross-over frequency ($\omega_c$), onset of interband transitions ($\omega_{int}$), and width of plasmon ($W_P$) of refractory ceramic materials.

| System | $a$ (Å) | $\omega_p$ (eV) | $\omega_c$ (eV) | $\omega_{int}$ (eV) | $W_P$ |
|---|---|---|---|---|---|
| TiC | 4.331 | 3.50 | 0.68 | < 1 eV | 35.2 |
| ZrC | 4.702 | 3.17 | 0.61 | < 1 eV | 36.2 |
| HfC | 4.642 | 3.52 | 0.73 | < 1 eV | 31.7 |
| TaC | 4.471 | 8.16 | 2.48 | < 1 eV | 6.57 |
| WC | 4.374 | 9.23 | 3.10 | <1 eV | 11.70 |
| TiN | 4.239 | 7.49 (6.69-8.7)[a] | 2.32 | 1.95 | 3.12 |
| ZrN | 4.578 | 8.63 (7.17-8.02)[a] | 2.86 | 2.60 | 1.09 |
| HfN | 4.521 | 9.10 (7.32-8.63)[a] | 3.25 | 3.11 | 1.24 |
| TaN | 4.401 | 10.68 (9.45)[a] | 3.94 | 3.41 | 3.15 |
| WN | 4.337 | 11.01 | 4.71 | 2.78 | 4.93 |

[a]Reference [10,31,42]

In short, the position of $E_F$ and the $p$ states of non-metallic elements influences the optical losses in the low energy region for these refractory materials. On the other hand, at higher energies (> 5-6 eV), the interband contribution for nitrides is stronger than for carbides. This is because of contributions from transitions along the BZ edges of X-W-L and K-W (see Fig. 1). Because our calculated spectra compare well with previously reported spectra derived from DFT calculations for TiN and ZrN,[31] we expect that our methodology will also remain valid for other compounds.

**Comparative performance with Au and Ag**

Next we discuss the comparative performances of all nitrides and one carbide (TaC) with Ag and Au.

As seen from Fig. 4, refractory materials, especially nitrides, exhibit better metallic behavior than Au and Ag in terms of real permittivity (smaller magnitude of real permittivity).[10] From the point of view of optical losses, the refractory materials perform better than Au in the visible region. On the other hand, Ag, which is an excellent low-loss material, shows better performance down to a wavelength of 450 nm. At shorter wavelengths, ZrN and HfN have lower losses than Ag. Overall, the low magnitude of the real permittivity (except ZrN and HfN) along with the low losses makes nitrides a good choice for plasmonic applications in the visible region.

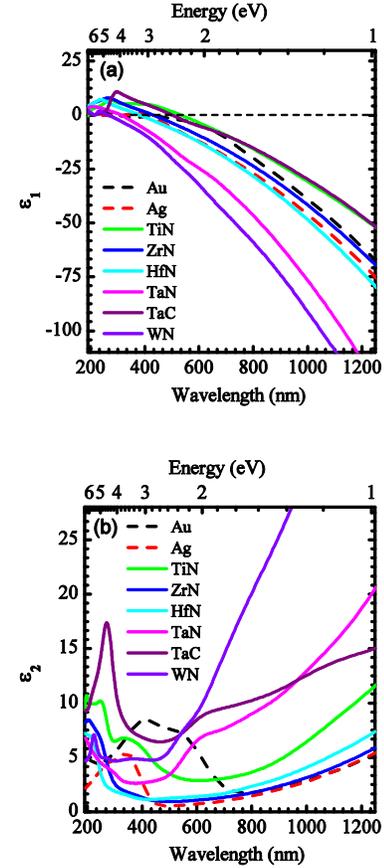

**Figure 4.** Comparison of calculated (a) real and (b) imaginary parts of the permittivites of selected refractory materials with Au and Ag as shown in dashed black and red lines, respectively.

To characterize the performance of selected refractory materials, we introduce two Q factors. One is for localized surface-plasmon resonance (LSPR), and the other is for surface plasmon polaritons (SPP), which are the localized and propagating types of surface plasmons, respectively. These dimensionless Q factors ($Q_{LSPR}$, and $Q_{SPP}$) are determined as follows:[18,43]

$$Q_{LSPR} = \frac{\omega \frac{d\varepsilon_1(\omega)}{d\omega}}{2\varepsilon_2(\omega)}, \qquad (9)$$

$$Q_{SPP} = \frac{\varepsilon_1(\omega)^2}{\varepsilon_2(\omega)}, \qquad (10)$$

These are shown in Fig. 5 along with Au and Ag Q factors (shown in dashed lines) for comparison.

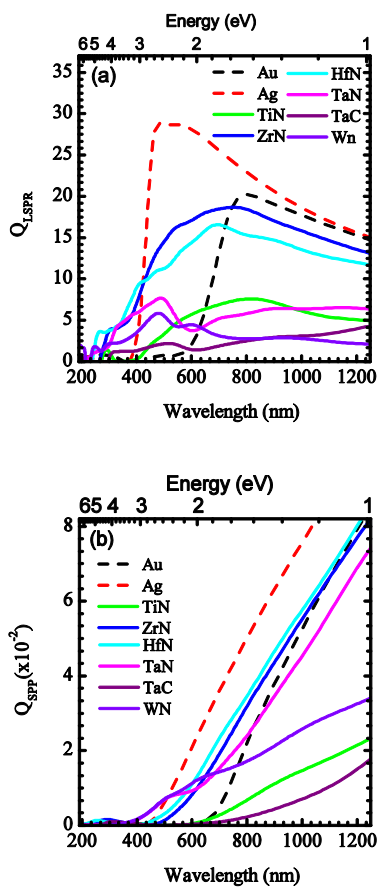

**Figure 5.** Calculated quality factors (a) $Q_{LSPR}$ and (b) $Q_{SPP}$, for selected refractory materials in comparison with those of Au and Ag.

As seen from Fig. 5, Ag outperforms the other materials. In the near-IR range, Au is the second best material. However, in the visible range HfN and ZrN are better than Au. Along with the chemical stability and cost of nitrides, the results suggest that HfN and ZrN are the materials of choice for a number of LSPR applications in the visible range, such as surface-enhanced Raman scattering, local field enhancement for sensing, metamaterials, and metasurfaces.[10] Surface plasmon waveguides and SPP sensing with HfN and ZrN are also promising applications

**Part B: Nanostructure properties calculations**

In this part the potential applications of refractory ceramics-based nanostructured materials are discussed. Among the applications of surface plasmons, LSPRs excited on plasmonic nanostructures have found numerous applications as discussed earlier. In addition to those related to resonance-enhanced radiative decay, it is worth mentioning that non-radiative decay is also enhanced simultaneously. To study LSPRs, the simplest case of a nanosphere embedded in an isotropic homogenous host material is considered. The scattering and absorption cross sections, $Q_{sca}$ and $Q_{abs}$, of these refractory ceramics are calculated within Mie theory. Using the DFT-calculated bulk dielectric functions and taking the particle radius as 50 nm and the host refractive index as 1.33, we plotted $Q_{sca}$ and $Q_{abs}$ of various refractory carbides and nitrides as shown in Fig. 6. One can see that ZrN has the sharpest and highest $Q_{sca}$ with a value of 6.3 at the wavelength of 559 nm (2.22 eV) followed closely by HfN with a value of 5.7 at 527 nm (2.35 eV). On the other hand, TiN has broad resonance with a maximum $Q_{sca}$ value of 2.39 at 654 nm (1.90 eV). TaN also shows a broadened resonance with a $Q_{sca}$ value of 4.32 at 458 nm (2.71 eV). However, a somewhat opposite trend is observed in the case of absorption efficiency of these materials; TiN shows the highest $Q_{abs}$ with a value of 3.33 at a wavelength of 637 nm (1.95 eV) followed by ZrN and HfN. Carbides, on the other hand, show very poor performance with regard to both $Q_{sca}$ and $Q_{abs}$. Two of the carbides, TaC and WC, show weak LSPRs. However, the peaks are broad because of large losses.

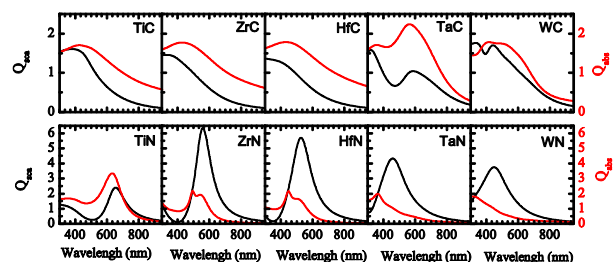

**Figure 6.** Calculated scattering ($Q_{sca}$) as black line and absorption ($Q_{abs}$) efficiencies as red line of single particles of carbides and nitrides based refractory materials.

**Comparison of the performance of refractory materials with Au and Ag**

Figure 7 compares the performance of selected nanostructured refractory materials with those of nanostructured Au and Au.

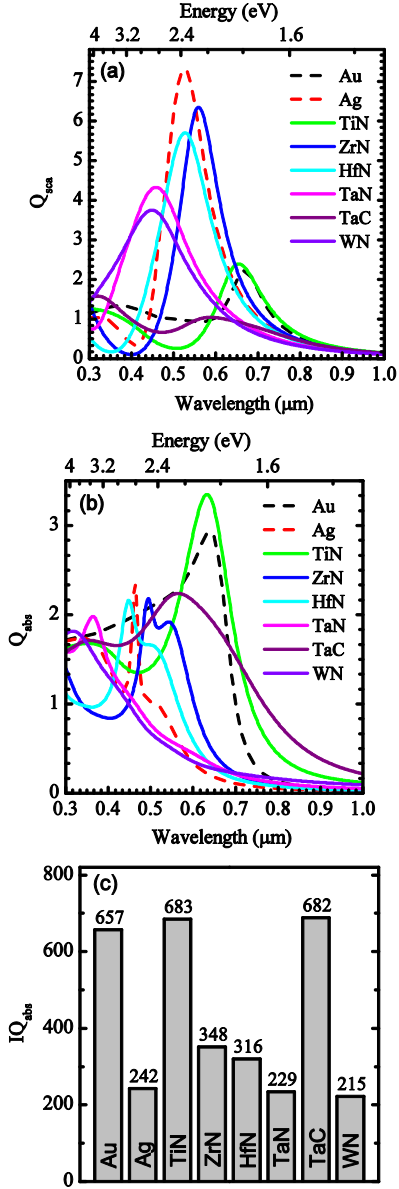

**Figure 7.** Calculated (a) $Q_{sca}$, (b) $Q_{abs}$ and (c) $IQ_{abs}$ for the selected refractory materials in comparison with those of Au and Ag.

As seen from Fig. 7a, among all the materials presented, Ag shows the best $Q_{sca}$ with a value of 7.31 at a wavelength of 526 nm (2.36 eV). However, $Q_{abs}$ of Ag (Fig. 7b) is low and comparable to that of ZrN and HfN. On the other hand, $Q_{abs}$ of TiN, with a maximum value of 3.33, is highest among all. Au, on the other hand shows an absorption efficiency of 2.94 at a wavelength of 640 nm (1.94 eV). Hence, it is worth mentioning that the TiN show better absorption efficiencies than Au. We also note that the peak position of TiN is in red part of the optical spectrum because of the smaller $\omega_p$. In addition to those factors related to either radiative or propagation efficiencies, we define another factor that characterizes non-radiative decay. To this end, we calculate the integral of $Q_{abs}$ ($IQ_{abs}$) in the range from 300 to 1300 nm, which corresponds to the major spectral composition of sunlight and also the transparent region of water.[44] $IQ_{abs}$ is calculated as follows:

$$IQ_{abs} = \int_{\lambda=300\ nm}^{\lambda=1300\ nm} Q_{abs}(\lambda) * E(\lambda)\ d(\lambda) \qquad (11)$$

where $E$ is normalized solar spectrum. Figure 7c shows the calculated $IQ_{abs}$ of selected refractory materials. The radius of the sphere (50 nm) and the host index of refraction (1.33) are the same as before. While the value of $Q_{abs}$ is important at single wavelength illumination, $IQ_{abs}$ is a useful measure for judging the degree of broadband absorption. As discussed previously,[45-47] aqueous solutions of nanoparticles are quite efficient at absorbing sunlight to heat water and generate vapor. When values of $IQ_{abs}$ are compared, as in Fig. 7c, a different order is seen for the materials under consideration. Because of broad resonances, the $IQ_{abs}$ of TiN and TaC are higher than that of Ag, and the $IQ_{abs}$ of TiN is even higher than that of Au. This comparison shows that TiN nanoparticles can be good broadband sunlight absorbers in the form of colloid suspensions or nanofluids.[44] It is important to note that 50-nm radius sphere is used to analyze the properties of these ceramic compounds. As in the case of metals,[18] changing the particle size can change the relative order of these materials.

## IV. CONCLUSIONS

First-principles DFT calculations have been performed to understand the electronic and optical properties of the refractory materials TiC, ZrC, HfC, TaC, WC, TiN, ZrN, HfN, TaN, and WN, for which not much experimental optical data are presently available. The main aim of this study is to systematically investigate how the electronic and optical properties vary by changing the non-metal species (carbon and nitrogen) as well as the metal species (Ti, Zr, Hf, Ta and W).

Our detailed investigation revealed that nitride based refractory compounds possess great potential for use as high-performance plasmonic compounds because of their highly metallic properties and low losses, especially for HfN and TaN in the visible region. On the other hand, carbides were found to be not suitable for plasmonic materials because of large losses from the visible to the near IR wavelength region. This loss generally originates from the interband optical transitions ($p \rightarrow d$) and from transitions from the Fermi surface to higher lying bands. These transitions occur in the low energy region below 1 eV in contrast to the nitride counterparts where the threshold energy for interband transitions is $\geq 2$ eV.

Analysis based on Mie theory suggests that TiN nanoparticles may be a good choice for broadband photothermal applications. This is because of the strong and broad resonance of TiN nanoparticles which makes them good sunlight absorbers in the form of nanofluids. TaC can also be used for photothermal applications because of its good broadband light absorption, which is comparable to that of Au. We also analyzed the LSPR and SPP properties of the refractory ceramics, and found that, within our theoretical framework, HfN and ZrN exhibit better performance than that of Au in the visible range, which is advantageous in terms of cost and their use in high-temperature environments[48]. Among the materials we have studied, TiN was found to be the best material for broadband photothermal applications, even better than Au or Ag, whereas ZrN and HfN are found to be candidates for low-loss, and thus narrow-band, plasmonic applications in the visible region.

Overall, our systematic studies covering ten different nitrides and carbides will serve as a guideline when choosing among these refractory ceramics for specific plasmonic applications.

**ACKNOWLEDGMENTS**
M.K. would like to thanks Dr. Byungki Ryu of Korea Electrotechnology Research Institute for useful discussion. This work was partly supported by the Core Research for Evolutional Science and Technology (CREST) program, the Japan Science and Technology Agency (JST) and the JSPS KAKENHI Grant Number 15K17447.